\documentclass[11pt]{article}
\usepackage{euscript}
\usepackage{amssymb}
\usepackage{amsfonts}
\usepackage{amsbsy}
\usepackage{epsfig}
\usepackage{amsthm}
\usepackage{amscd}
\usepackage{amstext}
\usepackage{verbatim}
\usepackage{amsmath}
\usepackage{hyperref}
\usepackage{cite}

\begin{document}

\begin{center}
{\Large \textbf{Collective diffusion and quantum chaos in holography }}

\vspace{1cm}

Shao-Feng Wu$^{1,2}$, Bin Wang$^{2,3}$, Xian-Hui Ge$^{1,2,4}$, Yu Tian$%
^{5,6,2} $

\vspace{1cm}

{\small \textit{$^{1}$Department of Physics, Shanghai University, Shanghai,
200444, China }}\\[0pt]
{\small \textit{$^{2}$Center for Gravitation and Cosmology, Yangzhou
University, Yangzhou 225009, China}}\\[0pt]
{\small \textit{$^{3}$Department of Physics and Astronomy, Shanghai Jiaotong
University, Shanghai, 200240, China}}\\[0pt]
{\small \textit{$^{4}$Department of Physics, University of California at San
Diego, California, 92093, USA}}\\[0pt]
{\small \textit{$^{5}$School of Physics, University of Chinese Academy of
Sciences, Beijing, 100049, China}}\\[0pt]
{\small \textit{$^{6}$Institute of Theoretical Physics, Chinese Academy of
Sciences, Beijing, 100190, China}}\\[0pt]
\vspace{0.5cm}

{\small \textit{sfwu@shu.edu.cn, wang\_b@sjtu.edu.cn, gexh@shu.edu.cn,
ytian@ucas.ac.cn}}
\end{center}

\vspace{1cm}

\begin{abstract}
We define\ a particular combination of charge and heat currents that is
decoupled with the heat current. This `heat-decoupled'\ (HD) current\ can be
transported by diffusion at long distances, when some thermo-electric
conductivities and susceptibilities satisfy a simple condition. Using the
diffusion condition together with the Kelvin formula, we show that the HD
diffusivity can be same as the charge diffusivity and also the heat
diffusivity. We illustrate that such mechanism is implemented in a strongly
coupled field theory, which is dual to a Lifshitz gravity with the dynamical
critical index $z=2$. In particular, it is exhibited that both charge and
heat diffusivities build the relationship to the quantum chaos. Moreover, we
study the HD diffusivity without imposing the diffusion condition. In some
homogeneous holographic lattices, it is found that the diffusivity/chaos
relation holds independently of any parameters, including the strength of
momentum relaxation, chemical potential, or temperature. We also show a
counter example of the relation and discuss its limited universality.
\end{abstract}

\pagebreak

\section{Introduction}

One of the most mysterious phenomenon in condensed matters is the ubiquitous
appearance of linear in temperature resistivity. In the materials with such
`strange-metal' hallmark, the quasiparticle picture is not applicable
because the resistivity can cross the Mott-Ioffe-Regel limit \cite%
{Hussey04,Gunnarsson03}. It has been long suggested \cite%
{Sachdevbook,ZaanenNature} that the transport in strange metals is
controlled by the `Planckian dissipation'---the temperature in units of time
through Planck's constant: $\tau _{\mathrm{P}}\sim \hbar /(k_{\mathrm{B}}T)$%
, but the explicit framework has not been built up.

In \cite{Hartnoll1405}, Hartnoll has pointed out that when the momentum
decays quickly, the collective diffusion of charge and energy controls the
strange-metal transport. Inspired by the putative bound on the ratio of
shear viscosity to entropy density \cite{KSS}, he proposed that the
eigenvalues of diffusive matrix are lower bounded by the Planckian
dissipation and Fermi velocity. Translating by the Einstein relation and
neglecting some thermoelectric effect, he argued that the saturation of the
diffusion bound may be responsible for the ubiquity of high-temperature
regimes in metals with $T$-linear resistivity.

The theory of incoherent metals is insightful, but there are several issues
which deserve to be explored. First, the diffusion bound can be violated in
various situations \cite{Hartnoll1612,Pakhira1409,Amoretti1411,LWJ1612}.
Second, some strange metals are relatively clean and their thermoelectric
effect may not be especially small. Third, the characteristic velocity $v_{%
\mathrm{P}}$ is taken as the Fermi velocity $v_{\mathrm{F}}$, which may not
be well defined in strongly coupled systems. To address the last issue, the
butterfly velocity $v_{\mathrm{B}}$ that quantifies the speed of chaos
propagation has been proposed as a natural replacement \cite%
{Blake1603,Blake1604}. Actually, an interesting relation has been found by
holography between the charge diffusivity and the quantum chaos which is
characterized by $v_{\mathrm{B}}$ and $\tau _{\mathrm{L}}$. Here $\tau _{%
\mathrm{L}}$ is the Lyapunov time that is expected to indicate the Planckian
dissipation in non-quasiparticle systems \cite%
{Larkin1696,Kitaev2014,Maldacena1503,Swingle1608,Patel1611}. However, in
addition to the non-universal prefactors that occur in special cases \cite%
{Lucas1608,Davison1612}, the application of charge diffusivity/chaos
relation is limited since the particle-hole symmetry must be imposed. As a
result, the relation does not appear in the normal state of cuprates. On the
other hand, the relation between the thermal diffusivity and chaos seems to
be more robust \cite%
{Blake1604,Blake1611,Niu1704,LWJ1705,Blake1705,Sachdev1611}, but it cannot
be directly translated into the statement of resistivity.

Nevertheless, by a recent local optical measurements of thermal diffusivity
on underdoped YBCO (an ortho-II YBa$_{2}$Cu$_{3}$O$_{6.60}$ and an ortho-III
YBa$_{2}$Cu$_{3}$O$_{6.75}$) crystals, it was found that the thermal
anisotropy is almost identical to the value of the electrical resistivity
anisotropy and starts to decrease sharply below the charge order transition
\cite{Zhang1610}. The experiment was interpreted that the non-quasiparticle
transport is dominated by the collective diffusion of electron-phonon
`soup'. In particular, it led to the conjecture that both charge and heat
diffusivities saturate the proposed bounds \cite{Zhang1610,Werman1705}.

To understand how the collective diffusion could be relevant to the
intrinsic mechanism for robust metallic transport that is complicated by the
extrinsic processes for momentum relaxation, it would be promising to search
hints in clean systems. Indeed, in a conformal field theory (CFT) with
charge doping, Davison, Gout\'{e}raux and Hartnoll (DGH) have isolated a
diffusive mode by hydrodynamics \cite{DGH,DG1505}, which is carried by a
particular combination of the charge and heat currents. The DGH mode can
likely be considered intrinsic, because the DGH current is decoupled with
momentum in the conformal fluid and the DGH conductivity is universal for
some clean holographic theories \cite{Jain}. However, the DGH mode has not
been studied when the translation symmetry is broken, partially because a
simple proposal to include slow momentum relaxation in hydrodynamics \cite%
{Hartnoll0706} is inconsistent with the holographic models \cite%
{DG1505,Blake1505} and the fast momentum relaxation invalidates the
hydrodynamics essentially.

In this paper, we will explore the collective diffusion of charge and energy
and its relationship to quantum chaos in terms of the gauge/gravity duality
with momentum relaxation. At the beginning, we will show that there is a
universal bound for thermo-electric transport. The bound is trivial, except
that its saturation indicates the decoupling between the heat current and a
particular combination of charge and heat currents. The `heat-decoupled'
(HD) current is nothing but the DGH\ current in most of homogeneous
holographic lattices. Without relying on the hydrodynamics, we will verify
that the HD\ current\ can be transported by diffusion at long distances,
when some thermo-electric conductivities and susceptibilities satisfy a
simple condition. Then we will study the HD mode by combining the diffusion
condition and the Kelvin formula \cite{Peterson1001}. Note that the Kelvin
formula arises if the static limit is taken before the thermodynamic limit
in the evaluation of Onsager coefficients \cite{Peterson1001}. It provides a
good approximate (sometimes even exact) expression of the thermopower in
various contexts including strongly correlated systems, such as the
fractional quantum Hall states \cite{Peterson1001}, high temperature
superconductors \cite{Peterson1001,Silk2009,Mravlje1504}, and the
homogeneous Sachdev-Ye-Kitaev model \cite{Davison1612}. As a result, we will
point out that not only the heat diffusivity but also the charge diffusivity
can be identified with the HD\ diffusivity. Furthermore, in a Lifshitz
gravity with the dynamical critical index $z=2$, we will illustrate that the
diffusion condition and the Kelvin formula can be realized. Meanwhile, the
HD\ diffusivity exhibits the relationship to the quantum chaos and is equal
to both charge and heat diffusivities.

In the thermo-electric systems that we care about, the fluctuations of
charge and energy are coupled and hence they do not satisfy separate
diffusion equations in general. Accordingly, the charge and energy
diffusivities do not correspond to any eigenvalues of the diffusivity
matrix. However, they still can be well defined (by their Einstein relation)
and would imply interesting physics. This inspires us to study the diffusion
constant of HD mode, but without imposing any diffusion condition. In other
words, the HD mode is no longer purely diffusive. We will exhibit that the
diffusivity/chaos relation is respected exactly in various homogeneous
holographic lattices. This is different with previous results in references,
where the diffusivity/chaos relation always requires certain limits on the
parameters, including the particle-hole symmetry \cite%
{Blake1603,Amoretti1411}, strong momentum relaxation \cite%
{Blake1604,Niu1704,LWJ1705}, or low temperature \cite{Blake1611,Blake1705}.
However, the relation is not universal and we will show a counter example.
The limited universality and implication will be discussed.

\section{Heat-decoupled current}

The thermo-electric transport is characterized by three conductivities:
electrical $\sigma $, thermal $\bar{\kappa}$, and thermoelectric $\alpha $%
\footnote{%
In this work, we focus on the dc conductivities.}. They reflect the response
of the charge and heat currents to small gradients of temperature and
chemical potential,%
\begin{equation}
\left(
\begin{array}{c}
J \\
J^{Q}%
\end{array}%
\right) =\left(
\begin{array}{cc}
\sigma & T\alpha \\
T\alpha & T\bar{\kappa}%
\end{array}%
\right) \left(
\begin{array}{c}
-\nabla \mu \\
-\nabla T/T%
\end{array}%
\right) .  \label{JJQ}
\end{equation}%
The DGH current is defined by a particular combination of charge and heat
currents \cite{DGH}%
\begin{equation}
J^{\mathrm{DGH}}\equiv \frac{sTJ-\rho J^{Q}}{sT+\rho \mu },
\end{equation}%
where $\rho $ is the charge density and $\mu $ is the chemical potential.
Now we construct a general linear combination but write it as the\ DGH-like
form for comparison%
\begin{equation}
\bar{J}^{\mathrm{DGH}}\equiv \frac{sTJ-\bar{\rho}J^{Q}}{sT+\bar{\rho}\mu }.
\end{equation}%
In the present, $\bar{\rho}$ is an arbitrary quantity with the dimension of
charge density. The current-current correlation can be calculated by
\begin{equation}
\bar{\sigma}_{\mathrm{DGH}}\equiv \left\langle \bar{J}^{\mathrm{DGH}}\bar{J}%
^{\mathrm{DGH}}\right\rangle =\frac{T(\bar{\rho}^{2}\bar{\kappa}%
+s^{2}T\sigma -2\bar{\rho}sT\alpha )}{\left( sT+\bar{\rho}\mu \right) ^{2}}.
\label{sDGH}
\end{equation}%
We reorganize the numerator of Eq. (\ref{sDGH}):%
\begin{eqnarray}
&&T(\bar{\rho}^{2}\bar{\kappa}+s^{2}T\sigma -2\bar{\rho}sT\alpha )  \notag \\
&=&\left( \sigma -\frac{T\alpha ^{2}}{\bar{\kappa}}\right) \left( sT\right)
^{2}+\frac{T}{\bar{\kappa}}\left( sT\alpha -\bar{\rho}\bar{\kappa}\right)
^{2}.  \label{Fenzi}
\end{eqnarray}%
In the dc limit, the last term in Eq. (\ref{Fenzi}) is non-negative, so we
immediately obtain a universal bound%
\begin{equation}
\bar{\sigma}_{\mathrm{DGH}}\geq \sigma _{0}\frac{\left( sT\right) ^{2}}{%
\left( sT+\bar{\rho}\mu \right) ^{2}},  \label{bound}
\end{equation}%
where $\sigma _{0}\equiv \sigma -T\alpha ^{2}/\bar{\kappa}$ denotes the
electric conductivity at zero heat current. The bound is almost trivial. The
only interesting point is that, due to
\begin{equation}
\left\langle \bar{J}^{\mathrm{DGH}}J^{Q}\right\rangle =\frac{\bar{\rho}\bar{%
\kappa}-sT\alpha }{\mu \left( sT+\bar{\rho}\mu \right) },
\end{equation}%
the bound is saturated when the general current is decoupled with the heat
current. Hereafter, we will focus on the HD current\footnote{%
Note that the current involves the two-point functions and hence cannot be
obtained directly by one variation of the generating functional.}%
\begin{equation}
J^{\mathrm{HD}}\equiv \left. \bar{J}^{\mathrm{DGH}}\right\vert _{\bar{\rho}%
=sT\alpha /\bar{\kappa}}=\frac{\bar{\kappa}J-\alpha J^{Q}}{\bar{\kappa}%
+\alpha \mu }.
\end{equation}%
One can find that the HD current is equal to the DGH current in most of
homogeneous holographic lattices which have\footnote{%
This relation was first noticed in \cite{Donos1406}.}%
\begin{equation}
\rho =sT\alpha /\bar{\kappa}.
\end{equation}%
However, this relation does not hold when the momentum relaxation is
inhomogeneous \cite{Donos1409} or involves some non-minimal coupling \cite%
{LWJ1602,LWJ1612,LWJ1705}. In these cases, the HD current is different with
the DGH current and might be viewed as an extension.

\section{Diffusive mode}

Suppose that the energy and the charge are conserved. The HD current
respects the continuous equation
\begin{equation}
\frac{\partial }{\partial t}\delta Q^{\mathrm{HD}}+\nabla \cdot J^{\mathrm{HD%
}}=0,  \label{cont eq}
\end{equation}%
and carries a particular combination of charge and energy%
\begin{equation}
\delta Q^{\mathrm{HD}}\equiv \delta \rho -\frac{\alpha }{\bar{\kappa}+\alpha
\mu }\delta \epsilon ,
\end{equation}%
where we denote $\epsilon $ as the energy density. Using the definition of
conductivities (\ref{JJQ}), the continuous equation (\ref{cont eq}), and the
thermodynamic identities%
\begin{eqnarray}
\nabla \rho &=&\chi \nabla \mu +\zeta \nabla T,  \notag \\
\nabla \epsilon &=&(T\zeta +\mu \chi )\nabla \mu +(\mu \zeta +c_{\mu
})\nabla T,  \label{drodE}
\end{eqnarray}%
with the susceptibilities defined by derivatives of pressure%
\begin{equation}
\chi \equiv \partial ^{2}p/\partial \mu ^{2},\;\zeta \equiv \partial
^{2}p/\partial \mu \partial T,\;c_{\mu }\equiv T\partial ^{2}p/\partial
T^{2},
\end{equation}%
one can prove at long distances%
\begin{equation}
\frac{\partial }{\partial t}\delta Q^{\mathrm{HD}}=D_{\mathrm{HD}}\left[
\nabla ^{2}\delta Q^{\mathrm{HD}}+C\nabla ^{2}\rho \right] ,  \label{diffeq2}
\end{equation}%
where we have defined%
\begin{eqnarray}
D_{\mathrm{HD}} &\equiv &\frac{\zeta }{\alpha }\frac{\bar{\kappa}\sigma
-T\alpha ^{2}}{c_{\mu }\chi -T\zeta ^{2}},  \label{D2} \\
C &\equiv &1-\frac{\alpha \mu }{\bar{\kappa}+\alpha \mu }\frac{c_{\mu
}+\zeta \mu }{\zeta \mu }.  \label{C2}
\end{eqnarray}%
Following Ref. \cite{Landau}, the diffusion equation%
\begin{equation}
\frac{\partial }{\partial t}\delta Q^{\mathrm{HD}}=D_{\mathrm{HD}}\nabla
^{2}\delta Q^{\mathrm{HD}}  \label{diff eq}
\end{equation}%
can be constructed, provided that $C=0$, which means%
\begin{equation}
\bar{\kappa}/c_{\mu }=\alpha /\zeta .  \label{cond3}
\end{equation}%
At this time, the diffusion constant can be written as%
\begin{equation}
\bar{D}_{\mathrm{HD}}\equiv \left. D_{\mathrm{HD}}\right\vert _{C=0}=\frac{%
\bar{\kappa}\sigma -T\alpha ^{2}}{\bar{\kappa}\chi -T\alpha \zeta }.
\label{D3}
\end{equation}

Some remarks are in order. First, the diffusion condition (\ref{cond3}) can
be understood as a special balance between the thermoelectric conductivity $%
\alpha $ and susceptibility $\zeta $. The thermoelectric balance is
reminiscent of the Kelvin formula which can be written as $\sigma /\chi
=\alpha /\zeta $. The Kelvin formula can be derived by requesting that the
density gradient vanishes \cite{Mravlje1504}. At that time, Eq. (\ref%
{diffeq2}) becomes the diffusion equation without requiring any additional
conditions. Second, usually the diffusion constant $D_{\mathrm{HD}}$ is not
equal to the charge diffusivity $D_{\mathrm{C}}\equiv \sigma /\chi $ or
thermal diffusivity $D_{\mathrm{T}}\equiv \kappa /c_{\rho }$, where $\kappa
\equiv \bar{\kappa}-T\alpha ^{2}/\sigma $ and $c_{\rho }\equiv c_{\mu
}-T\zeta ^{2}/\chi $. However, by recasting Eq. (\ref{D2}) as $D_{\mathrm{HD}%
}=D_{\mathrm{C}}D_{\mathrm{T}}\zeta /\alpha $ and invoking the Kelvin
formula $D_{\mathrm{C}}=\alpha /\zeta $, one can find $D_{\mathrm{HD}}=D_{%
\mathrm{T}}$. Furthermore, the combination of Eq. (\ref{cond3}) and the
Kelvin formula leads to $\bar{D}_{\mathrm{HD}}=D_{\mathrm{C}}=D_{\mathrm{T}}$%
. Third, $D_{\mathrm{HD}}=\sigma D_{\mathrm{T}}\zeta /\left( \chi \alpha
\right) $ apparently involves $\chi $ and $\zeta $. In the holographic
models, they depend on the full bulk geometry in general. However, they can
be cancelled in $D_{\mathrm{HD}}$ due to $\zeta /\chi =\left( \partial
s/\partial \rho \right) _{T}$, which is indeed related to the thermodynamics
of black holes \cite{Blake1611}. At last, it is instructive to clarify the
relation between the HD current with the diffusion condition and the
decoupled thermo-electric currents which are transported by diffusion \cite%
{DG1505,Ling1512}, see Appendix A. In the following, we will reveal a
general and exact relation between $D_{\mathrm{HD}}$ and $v_{\mathrm{B}%
}^{2}\tau _{\mathrm{L}}$ in various holographic models. In particular, the
diffusion condition (\ref{cond3}) and the Kelvin formula both hold in a
model of Lifshitz gravity with $z=2$, leading to $D_{\mathrm{C}}=D_{\mathrm{T%
}}=v_{\mathrm{B}}^{2}\tau _{\mathrm{L}}$.

\section{Holographic models}

\subsection{Einstein-Maxwell-axion}

A simple holographic framework with momentum relaxation was presented in
\cite{Andrade1311}. The model contains linear axions $\chi _{i}$ along
spatial directions. We consider the four-dimensional Einstein-Maxwell-axion
(EMA) theory,%
\begin{equation}
S_{\mathrm{bulk}}=\int d^{4}x\sqrt{-g}{\Big [}R+6-\frac{1}{4}F^{2}-\frac{1}{2%
}\sum\limits_{i=1}^{2}\left( \partial \chi _{i}\right) ^{2}{\Big ]}.
\end{equation}%
Here the AdS radius $L$ and the Newton constant $16\pi G_{N}$ are set to
unity. The equations of motion (EOM) derived from the action has an
isotropic background solution, in which $\chi _{i}=\beta x_{i}$ with the
disorder parameter $\beta $ \cite{Andrade1311}. Suppose that the horizon
locates at $r_{+}$. The Hawking temperature and entropy density are
\begin{equation}
T=\frac{1}{4\pi }\left( 3r_{+}-\frac{2\beta ^{2}+\rho ^{2}/r_{+}^{2}}{4r_{+}}%
\right) ,\;s=4\pi r_{+}^{2},  \label{Ts}
\end{equation}%
from which two thermodynamic response functions can be derived%
\begin{equation}
\left( \frac{\partial s}{\partial T}\right) _{\rho }=\frac{128\pi
^{2}r_{+}^{5}}{3\rho ^{2}+2\beta ^{2}r_{+}^{2}+12r_{+}^{4}},\;\left( \frac{%
\partial s}{\partial \rho }\right) _{T}=\frac{16\pi \rho r_{+}^{2}}{3\rho
^{2}+2\beta ^{2}r_{+}^{2}+12r_{+}^{4}}.  \label{dsdTEMA}
\end{equation}

By building up the conserved currents and invoking the sources that are
linear in time, Donos and Gauntlett derived the thermo-electric
conductivities analytically \cite{Donos1406}:%
\begin{equation}
\sigma =1+\frac{4\pi \rho ^{2}}{s\beta ^{2}},\;\alpha =\frac{4\pi \rho }{%
\beta ^{2}},\;\bar{\kappa}=\frac{4\pi sT}{\beta ^{2}}.  \label{sak}
\end{equation}%
Using Eq. (\ref{dsdTEMA}), Eq. (\ref{sak}), and $\zeta /\chi =\left(
\partial s/\partial \rho \right) _{T}$, the diffusion constant (\ref{D2})
can be calculated. The result is simple:%
\begin{equation}
D_{\mathrm{HD}}=\frac{1}{2r_{+}}.
\end{equation}

We need to study the quantum chaos. The onset of chaos is characterized by
the Lyapunov time $\tau _{\mathrm{L}}$ and the butterfly velocity $v_{%
\mathrm{B}}$. They can be calculated by constructing a shock wave near the
black hole \cite{Shenker1306,Roberts1409,Shenker1412,Swingle1603,Ling1610}.
For the EMA model, they are \cite{Blake1603,Blake1604}
\begin{equation}
\tau _{\mathrm{L}}=\frac{1}{2\pi T},\;v_{\mathrm{B}}^{2}=\frac{\pi T}{r_{+}}.
\label{vBEMA}
\end{equation}%
Remarkably, we have found an exact diffusivity/chaos relation%
\begin{equation}
D_{\mathrm{HD}}=v_{\mathrm{B}}^{2}\tau _{\mathrm{L}}.  \label{DHDEMA}
\end{equation}%
As a comparison, we write down the ratio between $D_{\mathrm{T}}$ and $v_{%
\mathrm{B}}^{2}\tau _{\mathrm{L}}$%
\begin{equation}
\frac{D_{\mathrm{T}}}{v_{\mathrm{B}}^{2}\tau _{\mathrm{L}}}=\frac{3\rho
^{2}+2r_{+}^{2}\beta ^{2}+12r_{+}^{4}}{4\left( \rho ^{2}+r_{+}^{2}\beta
^{2}\right) }.  \label{DTEMA}
\end{equation}

\subsection{Non-relativistic scaling}

We have interest on an Einstein-Maxwell-Axion-Dilaton (EMAD) theory studied
in \cite{Ge1606,Lu1608}%
\begin{equation}
S_{\mathrm{bulk}}=\int d^{4}x\sqrt{-g}\left[ R+V\left( \phi \right) -\frac{1%
}{2}\left( \partial \phi \right) ^{2}-\frac{1}{4}\sum\limits_{i=1}^{2}Z_{i}%
\left( \phi \right) F_{(i)}^{2}-\frac{1}{2}Y\left( \phi \right)
\sum\limits_{i=1}^{2}\left( \partial \chi _{i}\right) ^{2}\right] ,
\end{equation}%
where all the functions of the dilaton are assumed to have the exponential
form%
\begin{equation}
V\left( \phi \right) =-2\Lambda e^{\lambda _{0}\phi },\;Z_{1}\left( \phi
\right) =e^{\lambda _{1}\phi },\;Z_{2}\left( \phi \right) =e^{\lambda
_{2}\phi },\;Y\left( \phi \right) =e^{\lambda _{3}\phi },
\end{equation}%
with several parameters $\Lambda $ and $\lambda _{i}$\ ($i=0,1,2,3$). This
theory admits Lifshitz-like, hyperscaling violating, analytical black-brane
solutions. The line element is given by%
\begin{equation}
ds^{2}=r^{\theta }\left[ -r^{2z}f\left( r\right) dt^{2}+\frac{dr^{2}}{%
r^{2}f\left( r\right) }+r^{2}\left( dx^{2}+dy^{2}\right) \right] ,
\end{equation}%
where the blackness factor is%
\begin{equation}
f\left( r\right) =1-\frac{m}{r^{\theta +z+2}}+\frac{\beta ^{2}}{\left(
\theta +2\right) \left( z-2\right) }\frac{1}{r^{2z+\theta }}+\frac{q_{2}^{2}%
}{2\left( \theta +2\right) \left( z+\theta \right) }\frac{1}{r^{2(z+\theta
+1)}}.
\end{equation}%
We have denoted the dynamical critical index as $z$ and the hyperscaling
violating factor $\theta $. Besides the usual electromagnetic field, there
is an additional Maxwell field which is necessary for $z\neq 1$. Two Maxwell
fields have the charges $\left( q_{1},q_{2}\right) $, respectively. We will
impose the physical condition that the first U(1) current is vanishing. Then
the charge density is $\rho =q_{2}$. The parameters $\Lambda $, $\lambda
_{i} $, and $q_{1}$ are all determined by $z$ and $\theta $. For instance, $%
q_{1}=\sqrt{2\left( z-1\right) \left( z+\theta +2\right) }$. In \cite{Lu1608}%
, it has been pointed out that the black-brane solution is divergent when $%
z=2$, indicating the logarithmic behavior. The blackness factor then becomes%
\begin{equation}
f\left( r\right) =1-\frac{1}{r^{\theta +4}}\left( m+\frac{\beta ^{2}}{\theta
+2}\log r\right) +\frac{q_{2}^{2}}{2\left( \theta +2\right) ^{2}}\frac{1}{%
r^{2(\theta +3)}}.
\end{equation}%
Using the black-brane solutions, the temperature and entropy density can be
written as%
\begin{equation}
T=\frac{z+\theta +2}{4\pi }r_{+}^{z}-\frac{q_{2}^{2}+2\beta
^{2}r_{+}^{\theta +2}}{8\pi \left( \theta +2\right) r_{+}^{z+2\theta +2}}%
,\;s=4\pi r_{+}^{\theta +2},  \label{TsNR}
\end{equation}%
which is effective even for $z=2$. They can lead to two response functions%
\begin{eqnarray}
\left( \frac{\partial s}{\partial T}\right) _{\rho } &=&\frac{32\pi
^{2}r_{+}^{z+3\theta +4}(\theta +2)^{2}}{q_{2}^{2}\left( z+2\theta +2\right)
+2r_{+}^{\theta +2}\left[ \left( z+\theta \right) \beta
^{2}+r_{+}^{2z+\theta }z\left( \theta +2\right) \left( z+\theta +2\right) %
\right] },  \notag \\
\;\left( \frac{\partial s}{\partial \rho }\right) _{T} &=&\frac{8\pi
q_{2}r_{+}^{\theta +2}(\theta +2)}{q_{2}^{2}\left( z+2\theta +2\right)
+2r_{+}^{\theta +2}\left[ \left( z+\theta \right) \beta
^{2}+r_{+}^{2z+\theta }z\left( \theta +2\right) \left( z+\theta +2\right) %
\right] }.  \label{dsdTNR}
\end{eqnarray}%
From Ref. \cite{Ge1606}, the dc thermo-electric conductivities can be read
off,%
\begin{equation}
\sigma =r_{+}^{2z+\theta -2}+\frac{q_{2}^{2}r_{+}^{2z-\theta -2}}{\beta
^{2}r_{+}^{2-\theta }+q_{1}^{2}r_{+}^{2z+2}},\;\alpha =\frac{4\pi
q_{2}r_{+}^{2z}}{\beta ^{2}r_{+}^{2-\theta }+q_{1}^{2}r_{+}^{2z+2}},\;\bar{%
\kappa}=\frac{16\pi ^{2}Tr_{+}^{2z+\theta +2}}{\beta ^{2}r_{+}^{2-\theta
}+q_{1}^{2}r_{+}^{2z+2}}.  \label{NRDC}
\end{equation}

To obtain the quantities in chaos, we study a shock-wave metric%
\begin{equation}
ds^{2}=A(uv)dudv+V(uv)d\vec{x}^{2}-A(uv)\delta \left( u\right) h(t,\vec{x}%
)du^{2},
\end{equation}%
which is generated by releasing a particle from the boundary at $\vec{x}=0$.
From the EOM, one can obtain the solution of $h(t,\vec{x})$ at later time
and large distance. Then we can extract the Lyapunov time $\tau _{\mathrm{L}%
}=1/\left( 2\pi T\right) $ and the butterfly velocity $v_{\mathrm{B}}=2\pi
T/m$, with the screening length $m$ given by%
\begin{equation}
m^{2}=\frac{2}{A(0)}\left. \frac{\partial V(uv)}{\partial (uv)}\right\vert
_{u=0}.  \label{m2}
\end{equation}%
Equation (\ref{m2}) is same as the one in \cite{Blake1603}, indicating that
the additional Maxwell field does not change the form of the screening
length. Translating the Kruskal coordinates into Schwarzschild coordinates,
we have%
\begin{equation}
v_{\mathrm{B}}^{2}=\frac{2\pi Tr_{+}^{z-2}}{\theta +2}\mathcal{.}
\end{equation}

Using Eq. (\ref{dsdTNR}) and Eq. (\ref{NRDC}), one can calculate the
diffusion constant (\ref{D2})%
\begin{equation}
D_{\mathrm{HD}}=\frac{r_{+}^{z-2}}{\theta +2}.
\end{equation}%
Thus, we have the robust relation $D_{\mathrm{HD}}=v_{\mathrm{B}}^{2}\tau _{%
\mathrm{L}}$ for the generic $z$ and $\theta $. For comparison, the ratio
between $D_{\mathrm{T}}$ and $v_{\mathrm{B}}^{2}\tau _{\mathrm{L}}$ depends
on $z$ and $\theta $:
\begin{equation}
\frac{D_{\mathrm{T}}}{v_{\mathrm{B}}^{2}\tau _{\mathrm{L}}}=\frac{\left(
z+2\theta +2\right) q_{2}^{2}+2r_{+}^{\theta +2}\left[ \left( z+\theta
\right) \beta ^{2}+r_{+}^{2z+\theta }z\left( \theta +2\right) \left(
z+\theta +2\right) \right] }{2\left( \theta +2\right) \left[
q_{2}^{2}+r_{+}^{\theta +2}\beta ^{2}+2r_{+}^{2z+2\theta +2}\left(
z-1\right) \left( z+\theta +2\right) \right] }.
\end{equation}

Let's focus on the special case with $z=2$. From Eq. (\ref{dsdTNR}) and Eq. (%
\ref{NRDC}), one can see
\begin{equation}
\frac{\alpha }{\sigma }=\left( \frac{\partial s}{\partial \rho }\right) _{T},
\end{equation}%
which implies that the Kelvin formula $\sigma /\chi =\alpha /\zeta $ holds
and $D_{\mathrm{HD}}=D_{\mathrm{T}}$. Furthermore, we have interest on the
case with $\theta =0$. To be clear, we rewrite the temperature and entropy
density, and also invoke the chemical potential:%
\begin{equation}
T=\frac{1}{\pi }\left( r_{+}^{2}-\frac{2\beta ^{2}+\rho ^{2}/r_{+}^{2}}{%
16r_{+}^{2}}\right) ,\;s=4\pi r_{+}^{2},\;\mu =\frac{\rho }{2r_{+}^{2}}.
\label{TsNR2}
\end{equation}%
The last equation is necessary to calculate the quantities $\chi $, $\zeta $%
, and $c_{\mu }$. One can find that the diffusion condition (\ref{cond3}) is
obeyed. Collecting all the results in the exceptional case with $z=2$ and $%
\theta =0$, we have a series of equalities%
\begin{equation}
D_{\mathrm{HD}}=\bar{D}_{\mathrm{HD}}=D_{\mathrm{C}}=D_{\mathrm{T}}=v_{%
\mathrm{B}}^{2}\tau _{\mathrm{L}}.
\end{equation}

\subsection{High curvature}

The higher derivative corrections appear generally in any quantum gravity
theory from quantum or stringy effects. These corrections may be holographic
dual to $1/N$ or $1/\lambda $ corrections in some gauge theories, allowing
independent values of two central charges $a$ and $c$. This is in contrast
to the standard $\mathcal{N}$=4 super Yang-Mills theory where $a=c$, hence
likely underpinning the violation of viscosity bound. Actually, the
Gauss-Bonnet (GB) correction has been treated as a dangerous source of
violation for the feature that is universal in Einstein gravity \cite%
{Kats0712,Liu0802}.

Consider the GB correction to the EMA theory, that is%
\begin{eqnarray}
S_{\mathrm{bulk}} &=&\int d^{5}x\sqrt{-g}{\Big [}R+12-\frac{1}{4}F^{\mu \nu
}F_{\mu \nu }-\frac{1}{2}\sum\limits_{i=1}^{3}\left( \partial \chi
_{i}\right) ^{2}  \notag \\
&&+\frac{\tilde{\alpha}}{2}\left( R^{2}-4R^{\mu \nu }R_{\mu \nu }+R_{\mu \nu
\lambda \rho }R^{\mu \nu \lambda \rho }\right) {\Big ]},
\end{eqnarray}%
where $\tilde{\alpha}$ is the GB coupling constant. The black-brane solution
with $\chi _{i}=\beta x_{i}$ has been derived in \cite{Ge1411GB}. The
temperature and entropy density can be written as%
\begin{equation}
T=\frac{L_{\mathrm{eff}}}{\pi }\left( r_{+}-\frac{\beta ^{2}}{8r_{+}}-\frac{%
\rho ^{2}}{24r_{+}^{5}}\right) ,\;s=4\pi r_{+}^{3}.  \label{GBTsro}
\end{equation}%
Here $L_{\mathrm{eff}}^{2}=\frac{1+\sqrt{1-4\tilde{\alpha}}}{2}$ is the
square of the effective AdS radius. Taking the derivatives of the entropy
density, the two response functions read:%
\begin{equation}
\left( \frac{\partial s}{\partial T}\right) _{\rho }=\frac{1}{L_{\mathrm{eff}%
}}\frac{288\pi ^{2}r_{+}^{8}}{5\rho ^{2}+3\beta ^{2}r_{+}^{4}+24r_{+}^{6}}%
,\;\left( \frac{\partial s}{\partial \rho }\right) _{T}=\frac{24\pi \rho
r_{+}^{3}}{5\rho ^{2}+3\beta ^{2}r_{+}^{4}+24r_{+}^{6}}.  \label{dsdTGB}
\end{equation}%
We also need the thermo-electric conductivities \cite{Ge1411GB}
\begin{equation}
\sigma =r_{+}+\frac{4\pi \rho ^{2}}{s\beta ^{2}},\;\alpha =\frac{4\pi \rho }{%
\beta ^{2}},\;\bar{\kappa}=\frac{4\pi sT}{\beta ^{2}}.  \label{GBsigmaak}
\end{equation}%
Using Eq. (\ref{dsdTGB}) and Eq. (\ref{GBsigmaak}), the diffusion constant (%
\ref{D2}) can be calculated as%
\begin{equation}
D_{\mathrm{HD}}=\frac{L_{\mathrm{eff}}}{3r_{+}}.
\end{equation}

The butterfly effect for GB gravity has been studied in \cite{Roberts1409}.
We will repeat the calculation but involve the linear axions. Release a
particle from the boundary and consider the five-dimensional shock-wave
metric%
\begin{equation}
ds^{2}=A(uv)dudv+V(uv)d\vec{x}^{2}-A(uv)\delta \left( u\right) h(t,\vec{x}%
)du^{2}.
\end{equation}%
Using the EOM, one can derive the solution of $h(t,\vec{x})$ at later time
and large distance, which in turn gives the Lyapunov time $\tau _{\mathrm{L}%
}=1/\left( 2\pi T\right) $ and the butterfly velocity $v_{\mathrm{B}}=2\pi
T/m$, with the screening length $m$ given by%
\begin{equation}
m^{2}=\frac{3}{A(0)}\left. \frac{\partial V(uv)}{\partial (uv)}\right\vert
_{u=0}.  \label{sl}
\end{equation}%
Under the Schwarzschild coordinates, one can read%
\begin{equation}
v_{\mathrm{B}}^{2}=2\pi T\frac{L_{\mathrm{eff}}}{3r_{+}}\mathcal{.}
\label{vB2GB}
\end{equation}%
Then we build the diffusivity/chaos relation $D_{\mathrm{HD}}=v_{\mathrm{B}%
}^{2}\tau _{\mathrm{L}}$ again. The ratio between $D_{\mathrm{T}}$ and $v_{%
\mathrm{B}}^{2}\tau _{\mathrm{L}}$ can be given by%
\begin{equation}
\frac{D_{\mathrm{T}}}{v_{\mathrm{B}}^{2}\tau _{\mathrm{L}}}=\frac{5\rho
^{2}+3r_{+}^{4}\beta ^{2}+24r_{+}^{6}}{6\left( \rho ^{2}+r_{+}^{4}\beta
^{2}\right) }.
\end{equation}

\subsection{Anisotropy}

Spatially anisotropic black-brane is the first example that yields the
violation of viscosity bound in Einstein gravity \cite{Rebhan1110}. A
careful examination of the diffusivity/chaos relation in such anisotropic
model is worthwhile. In \cite{Ge1404,Ge1412}, a R-charged version of the
spatially anisotropic black-brane solution was derived via nonlinear
Kaluza-Klein reduction of type-IIB supergravity to five dimensions, which
leads to the presence of an Abelian field in the action. The introduction of
the $U(1)$ gauge field breaks the $SO(6)$ symmetry and thus leads to the
excitations of the Kaluza-Klein modes. Consider the reduced theory that is
described by a five-dimensional EMAD action%
\begin{equation}
S_{\mathrm{bulk}}=\int d^{5}x\sqrt{-g}\left[ R+12-\frac{1}{4}F^{\mu \nu
}F_{\mu \nu }-\frac{1}{2}\left( \partial \phi \right) ^{2}-\frac{1}{2}%
e^{2\phi }\left( \partial \chi \right) ^{2}\right] .
\end{equation}%
There is an anisotropic solution%
\begin{equation}
ds^{2}=-h(r)dt^{2}+\frac{1}{f(r)}dr^{2}+g_{11}(r)dx_{1}^{2}+g_{33}(r)\left(
dx_{2}^{2}+dx_{3}^{2}\right)
\end{equation}%
\begin{equation}
A=A_{t}(r)dt,\;\phi =\phi (r),\;\chi =\beta x_{1},  \label{Animetric}
\end{equation}%
which has been found numerically in \cite{Ge1404}. Note that the strength of
anisotropy and disorder are both characterized by the single parameter $%
\beta $. The analytical solution with small $\beta $ has also been obtained,
see Appendix D in \cite{Ge1404}. We write them as follows:%
\begin{eqnarray}
h(r) &=&\frac{1}{u^{2}}e^{-\frac{\phi (u)}{2}}\mathcal{B}(u)\mathcal{F}%
(u),\;f(r)=\frac{1}{u^{2}}e^{\frac{\phi (u)}{2}}\mathcal{F}(u),  \notag \\
g_{11}(r) &=&\frac{1}{u^{2}}e^{-\frac{3\phi (u)}{2}},\;g_{33}(r)=\frac{1}{%
u^{2}}e^{-\frac{\phi (u)}{2}},  \label{gxymetric}
\end{eqnarray}%
where $u\equiv 1/r$. The three functions $\left( \mathcal{F},\mathcal{B}%
,\phi \right) $ are given by%
\begin{eqnarray}
\mathcal{F}(u) &=&1-\frac{u^{4}}{u_{+}^{4}}-\frac{1}{12}\rho
^{2}u^{4}u_{+}^{2}+\frac{\rho ^{2}u^{6}}{12}+\beta ^{2}\mathcal{F}_{2}(u)+%
\mathcal{O(}\beta ^{4}\mathcal{),}  \notag \\
\mathcal{B}(u) &=&1+\beta ^{2}\mathcal{B}_{2}(u)+\mathcal{O(}\beta ^{4}%
\mathcal{),}  \notag \\
\phi (u) &=&\beta ^{2}\phi _{2}(u)+\mathcal{O(}\beta ^{4}\mathcal{),}
\label{gxymetric2}
\end{eqnarray}%
where $\left( \mathcal{F}_{2},\mathcal{B}_{2},\phi _{2}\right) $ can be read
off from Eq. (135) in \cite{Ge1404} with the replacement for our convention%
\begin{equation}
\left( a,u_{\mathrm{H}},q^{2}\right) \rightarrow \left( \beta ,u_{+},\frac{%
\rho ^{2}u_{+}^{6}}{12}\right) .
\end{equation}%
To be simple, we will use the analytical solution to study the diffusion and
chaos, instead of the numerical solution. The two thermodynamic quantities
are%
\begin{equation}
T=\frac{\sqrt{\mathcal{B}(u_{+})}\mathcal{F}^{\prime }(u_{+})}{4\pi }%
,\;s=4\pi u_{+}^{-3}e^{-\frac{5}{4}\phi (u_{+})}.
\end{equation}%
We need the thermo-electric conductivities, which can be found in \cite%
{Ge1412}:%
\begin{equation}
\sigma =\frac{g_{33}(u_{+})}{\sqrt{g_{11}(u_{+})}}+\frac{4\pi \rho ^{2}}{%
e^{2\phi (r_{+})}\beta ^{2}s},\;\alpha =\frac{4\pi \rho }{e^{2\phi
(r_{+})}\beta ^{2}},\;\bar{\kappa}=\frac{4\pi sT}{e^{2\phi (r_{+})}\beta ^{2}%
}.  \label{gxysak}
\end{equation}

We will study the anisotropic chaos. The present model is the
five-dimensional anisotropic model with an axion. Note that the butterfly
velocity in the four-dimensional model with anisotropic lattices has been
studied recently in \cite{Ling1610}. Consider a shock-wave metric%
\begin{eqnarray}
ds^{2} &=&A(uv)dudv+V_{1}(uv)dx_{1}^{2}+V_{3}(uv)\left(
dx_{2}^{2}+dx_{3}^{2}\right)  \notag \\
&&-A(uv)\delta \left( u\right) h(t,\vec{x})du^{2}.
\end{eqnarray}%
One can find that the previous form of Lyapunov time $\tau _{\mathrm{L}%
}=1/\left( 2\pi T\right) $ and butterfly velocity $v_{\mathrm{B}}^{2}=2\pi
T/m$ is not changed, but the screening length is affected by anisotropy:
\begin{equation}
m^{2}=\frac{V_{1}(0)}{A(0)}\left. \sum_{i=1}^{3}\frac{V_{i}^{\prime }(0)}{%
V_{i}(0)}\right\vert _{u=0}.
\end{equation}

\begin{figure}[tbp]
\centerline{
\includegraphics[width=.6\textwidth]{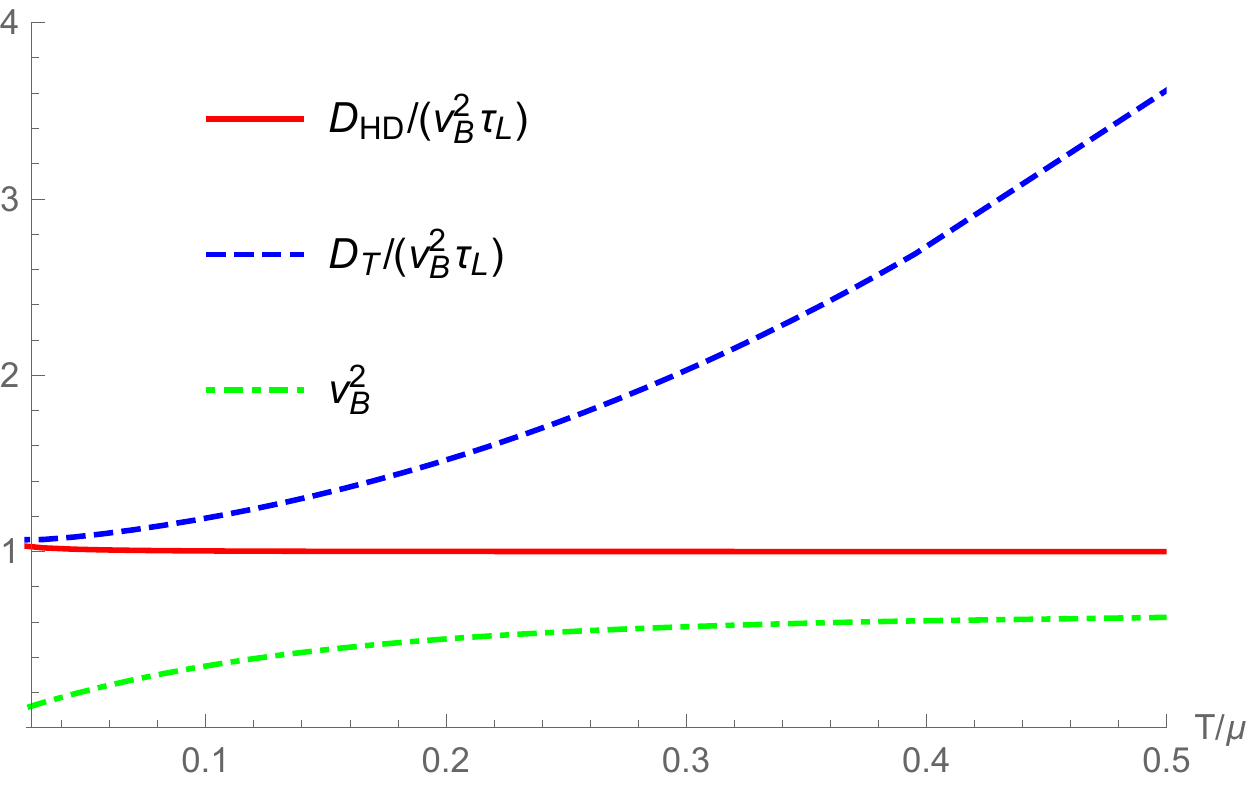}}
\caption{Diffusivty/chaos relation in the anisotropic holographic model with
$\protect\beta /\protect\mu =0.1$.}
\label{DHD}
\end{figure}

Using the analytical expression of the background solution and
conductivities, we display in Figure \ref{DHD} the temperature dependence of
the butterfly velocity and two diffusivity/chaos relation for comparison.
Typically, we fix the parameter $\beta /\mu =0.1$ and cut off the regime $%
T/\mu \ll 0.1$ in which $\beta \gg T$ and hence the analytical solution is
not reliable. Other values of the parameter do not lead to the qualitative
difference. As a result, one can see $D_{\mathrm{HD}}=v_{\mathrm{B}}^{2}\tau
_{\mathrm{L}}$.

\subsection{Non-minimal coupling}

In all aforementioned models, the translation-symmetry breaking sector is
minimally coupled to the gravitational and electromagnetic sectors. There
are novel models which involve the non-minimal coupling between the Maxwell
term and the axions \cite{LWJ1602,LWJ1612}. We will focus on one of these
models, i.e. the model 1 in \cite{LWJ1602}. Compared with others, its
conductivities are more trivial, leading to the situation with $J^{\mathrm{HD%
}}\neq J^{\mathrm{DGH}}$. Actually, this model is so distinctive that it
breaks various bounds on the viscosity \cite{KSS}, electric conductivity
\cite{Sachdev1507} and charge diffusivity \cite{Hartnoll1405}. The action is
given by
\begin{equation}
S_{\mathrm{bulk}}=\int d^{4}x\sqrt{-g}\left( R+6-\frac{1}{4}F^{2}-\frac{1}{4}%
\mathcal{J}\mathrm{Tr}\left[ \mathcal{X}F^{2}\right] -\mathrm{Tr}\left[
\mathcal{X}\right] \right) ,
\end{equation}%
where $\mathcal{J}$ is the coupling constant and%
\begin{equation}
\mathcal{X}_{\;\nu }^{\mu }=\frac{1}{2}\sum\limits_{i=1}^{2}\partial ^{\mu
}\chi _{i}\partial _{\nu }\chi _{i}.
\end{equation}%
The background solution is same as that in the EMA model. Thus, the
thermodynamical quantities (\ref{Ts}) are still applicable. From Ref. \cite%
{LWJ1612}, one can read off the thermoelectric conductivities%
\begin{eqnarray}
\sigma &=&\frac{\left( s-\pi \mathcal{J}\beta ^{2}\right) \left( 4\pi \rho
^{2}+s\beta ^{2}\right) }{\beta ^{2}\left( s^{2}+4\pi ^{2}\mathcal{J}\rho
^{2}\right) },\;\alpha =\frac{4\pi \rho s\left( s-\pi \mathcal{J}\beta
^{2}\right) }{\beta ^{2}\left( s^{2}+4\pi ^{2}\mathcal{J}\rho ^{2}\right) },
\notag \\
\bar{\kappa} &=&\frac{4\pi Ts^{3}}{\beta ^{2}\left( s^{2}+4\pi ^{2}\mathcal{J%
}\rho ^{2}\right) }.  \label{sakLWJ}
\end{eqnarray}%
Note that the HD current can be determined by the quantity $\bar{\rho}%
=sT\alpha /\bar{\kappa}=\rho (1-\pi \mathcal{J}\beta ^{2}/s)$.

Using Eq. (\ref{dsdTEMA}) and Eq. (\ref{sakLWJ}), we find that the $D_{%
\mathrm{HD}}$ is not changed by the non-minimal coupling, that is, $D_{%
\mathrm{HD}}=1/\left( 2r_{+}\right) $. Since the butterfly velocity is also
same as the one of EMA theory \cite{LWJ1612}, the relation $D_{\mathrm{HD}%
}=v_{\mathrm{B}}^{2}\tau _{\mathrm{L}}$ remains valid. The non-minimal
coupling also does not affect the ratio (\ref{DTEMA}).

\subsection{Nonlinear electromagnetism}

Born-Infeld theory is the simplest nonlinear generalization of Maxwell
electromagnetism \cite{BI1934}. A recent review on Born-Infeld gravity can
be found in \cite{Jimenez1704}. Adding two linear axions in the Born-Infeld
theory gives rise to the action%
\begin{equation}
S_{\mathrm{bulk}}=\int d^{4}x\sqrt{-g}\left[ R+6+\frac{1}{b}\left( 1-\sqrt{1+%
\frac{b}{2}F^{\mu \nu }F_{\mu \nu }}\right) -\frac{1}{2}\sum%
\limits_{i=1}^{2}\left( \partial \chi _{i}\right) ^{2}\right] ,
\label{BIaction}
\end{equation}%
where $b$ is the Born-Infeld parameter. Without the axions, the black-brane
solution that is asymptotically AdS can be found in \cite{Cai04,Dey04}. In
Appendix B, we extend the solution to involve the axions. Using the metric
ansatz (\ref{BIansatz2}) and the blackness factor (\ref{blacknessBI}), the
Hawking temperature and the entropy density can be obtained%
\begin{equation}
T=\frac{1}{4\pi }\left\{ 3r_{+}-\frac{\beta ^{2}}{2r_{+}}+\frac{r_{+}}{2b}%
\left( 1-Z_{eff}\right) \right\} ,\;s=4\pi r_{+}^{2},  \label{TsBI}
\end{equation}%
where $Z_{eff}$ denotes the effective electromagnetic coupling on the
horizon, see Eq. (\ref{BIAt}). From Eq. (\ref{TsBI}), one can get the
derivatives%
\begin{eqnarray}
\left( \frac{\partial s}{\partial T}\right) _{\rho } &=&\frac{64\pi
^{2}Z_{eff}r_{+}^{5}}{b\left[ \rho ^{2}+Z_{eff}r_{+}^{2}\left(
6r_{+}^{2}+\beta ^{2}\right) \right] -\left( 1-Z_{eff}\right) r_{+}^{4}},
\notag \\
\left( \frac{\partial s}{\partial \rho }\right) _{T} &=&\frac{8\pi \rho
r_{+}^{2}b}{b\left[ \rho ^{2}+Z_{eff}r_{+}^{2}\left( 6r_{+}^{2}+\beta
^{2}\right) \right] -\left( 1-Z_{eff}\right) r_{+}^{4}}.  \label{dsdTBI}
\end{eqnarray}%
In Appendix B, we have calculated the dc thermo-electric conductivities by
the Donos-Gauntlett method \cite{Donos1406}:%
\begin{equation}
\sigma =Z_{eff}+\frac{4\pi \rho ^{2}}{\beta ^{2}s},\;\alpha =\frac{4\pi \rho
}{\beta ^{2}},\;\bar{\kappa}=\frac{4\pi sT}{\beta ^{2}}.
\end{equation}%
Using the derivatives of the entropy and the thermo-electric conductivities,
we obtain $D_{\mathrm{HD}}=1/\left( 2r_{+}\right) $ again. Performing a
similar calculation as in the EMA model, we find that the Lyapunov time $%
\tau _{\mathrm{L}}=1/\left( 2\pi T\right) $ and the butterfly velocity $v_{%
\mathrm{B}}^{2}=\pi T/r_{+}$ are not changed by the nonlinear
electromagnetism. Then the relation $D_{\mathrm{HD}}=v_{\mathrm{B}}^{2}\tau
_{\mathrm{L}}$ is not changed either. On the contrary, the ratio between $D_{%
\mathrm{T}}$ and $v_{\mathrm{B}}^{2}\tau _{\mathrm{L}}$ is changed:%
\begin{equation}
\frac{D_{\mathrm{T}}}{v_{\mathrm{B}}^{2}\tau _{\mathrm{L}}}=\frac{b\left[
\rho ^{2}+Z_{eff}r_{+}^{2}\left( 6r_{+}^{2}+\beta ^{2}\right) \right]
-\left( 1-Z_{eff}\right) r_{+}^{4}}{2b\left( \rho ^{2}+Z_{eff}r_{+}^{2}\beta
^{2}\right) }.
\end{equation}

\subsection{Massive gravity}

The massive gravity with a reference metric is a well-known gravitational
model which breaks the diffeomorphism symmetry explicitly \cite%
{Pauli1939,Veltman1970,deRham2010}. Recently, it is applied to the
gauge/gravity duality, where the reference metric imitates the mean-field
disorder \cite{Vegh2013}. The action of the model is%
\begin{equation}
S_{\mathrm{bulk}}=\int d^{4}x\sqrt{-g}\left[ R+6-\frac{1}{4}F^{\mu \nu
}F_{\mu \nu }-\frac{1}{2}\beta ^{2}\left( \left( \mathrm{Tr}\mathcal{K}%
\right) ^{2}-\mathrm{Tr}\mathcal{K}^{2}\right) \right] ,
\end{equation}%
where the matrix $\mathcal{K}$ is defined by a matrix square root $\mathcal{K%
}_{\;\nu }^{\mu }=\sqrt{g^{\mu \lambda }b_{\lambda \nu }}$, with $b_{\mu \nu
}=\mathrm{diag}(0,0,1,1)$. It has been found that this model admits an
analytical black-brane solution, from which the thermodynamic quantities can
be obtained \cite{Vegh2013,Amoretti1406}. The analytical dc thermoelectric
conductivities have been calculated in \cite{Amoretti1407}. One can find
that they are all same to the ones of EMA models, see Eq. (\ref{Ts}) and Eq.
(\ref{sak}). The chaos in the massive gravity has not been studied before.
But we have checked that the Lyapunov time and the butterfly velocity are
the same to Eq. (\ref{vBEMA}). As a result, the massive gravity and the EMA
model share the same diffusivity/chaos relation.

\section{Discussion}

We have studied the collective diffusion of charge and energy in the
strongly coupled systems with finite density. Our focus is a particular
combination of charge and heat currents that decouples with the heat current
and can be reduced to the DGH current in most of homogeneous holographic
lattices. We derived the condition under which the HD current can be
transported by diffusion. Using the diffusion condition, Kelvin formula, and
holography, we have provided a mechanism for the relation between the
diffusion and chaos $D_{\mathrm{C}}\sim D_{\mathrm{T}}\sim v_{\mathrm{B}%
}^{2}\tau _{\mathrm{L}}$. This might be interesting because the evidence of
the relation $D_{\mathrm{C}}\sim D_{\mathrm{T}}$ in the normal state of the
cuprates was presented by a recent experiment and it was conjectured that
both charge and heat diffusivities saturate the Planckian bound \cite%
{Zhang1610}. Note that the relation $D_{\mathrm{C}}\sim D_{\mathrm{T}}\sim
v_{\mathrm{B}}^{2}\tau _{\mathrm{L}}$ was previously obtained in the EMA
model when the momentum relaxation is very strong and in the generic
holographic scaling geometries with a particle-hole symmetry \cite{Blake1604}%
. Our mechanism does not require the incoherent limit and zero-density
limit. Instead, the balance between the thermoelectric effect $\alpha $ and $%
\zeta $ is important, which is reflected in the diffusion condition $\bar{%
\kappa}/c_{\mu }=\alpha /\zeta $ and the Kelvin formula $\sigma /\chi
=\alpha /\zeta $.

Moreover, we have found a new diffusivity/chaos relation $D_{\mathrm{HD}%
}\sim v_{\mathrm{B}}^{2}\tau _{\mathrm{L}}$ that is universal in some
holographic models. Most of these models are based on the EMA theory,
involving various corrections: the non-relativistic scaling, the high
curvature, the anisotropy, the non-minimal coupling, and the nonlinear
electromagnetism. The theory of massive gravity is also taken into account.
Nevertheless, it is not very difficult to find a counter example, see
Appendix C. We attempt to understand the limited universality and
implication as follows.

(i) There are two reasons at least that the heat diffusivity can be
connected to the quantum chaos in a wide class of theories\footnote{%
In \cite{Sachdev1611}, the intuitive picture of the connection between chaos
and energy transport is depicted by recognizing that the quantum chaos is
linked to the loss of phase coherence and in turn the energy fluctuations.}.
On the one hand, when the chemical potential is nonvanishing, the
open-circuit heat conductivity $\kappa $ is finite in the translationally
invariant limit \cite{Blake1705}. Thus, it may be intrinsic in the sense
that it is not sensitive to irrelevant deformations that dissipate the
momentum \cite{Hartnoll1304}. On the other hand, from the perspective of
holography, $D_{\mathrm{T}}=\kappa /c_{\rho }$ can be determined solely by
the near-horizon physics \cite{Blake1604,Blake1611,Blake1705}. Thus, the
universal features may be emergent due to the similarity of all horizons.
Interestingly, $D_{\mathrm{HD}}=D_{\mathrm{C}}D_{\mathrm{T}}\zeta /\alpha $
is also finite in the translationally invariant limit. This can be verified
readily in terms of the hydrodynamics for clean systems with charge doping:
using Eq. (2) in Ref. \cite{DGH}, one can read $\sigma /\alpha =\rho /s$ in
the dc limit and thereby $D_{\mathrm{HD}}=D_{\mathrm{T}}\zeta \rho /\left(
s\chi \right) $ is finite. Moreover, as we pointed out in Sec. 3, the
quantities $\chi $ and $\zeta $ that depend on the full bulk geometry can be
cancelled in $D_{\mathrm{HD}}$. As a result, the HD diffusivity is also
determined solely by the near-horizon physics.

(ii) Based on the EMAD theory, it was found that the relation $D_{\mathrm{T}%
}\sim v_{\mathrm{B}}^{2}\tau _{\mathrm{L}}$ and the Kelvin formula are both
respected in the holographic models that flow to\ the $AdS_{2}\times R^{d}$
fixed points in the IR \cite{Blake1611}. Immediately, this leads to $D_{%
\mathrm{HD}}=D_{\mathrm{T}}\sim v_{\mathrm{B}}^{2}\tau _{\mathrm{L}}$ at low
temperatures. One can find that all of the isotropic holographic models that
we have studied have the $AdS_{2}$ horizon\footnote{%
It has been argued that in the zero temperature limit, the anisotropic
black-brane solution has a Lifshitz-like region in the IR \cite{Mateos1106}.
However, our analytic solution cannot achieve the zero temperature.},
including the counter example of the relation $D_{\mathrm{HD}}\sim v_{%
\mathrm{B}}^{2}\tau _{\mathrm{L}}$ given in Appendix C. This implies that
the $AdS_{2}$ horizon might be conducive to the existence of the relation $%
D_{\mathrm{HD}}\sim v_{\mathrm{B}}^{2}\tau _{\mathrm{L}}$, but it is not a
sufficient condition.

(iii) In the various models studied in the main text, the relation $D_{%
\mathrm{HD}}\sim v_{\mathrm{B}}^{2}\tau _{\mathrm{L}}$ exactly holds
independently of the temperature, chemical potential, and strength of
momentum relaxation. In particular, the diffusion condition is not satisfied
in these holographic models, except the Lifshitz gravity with $z=2$. This
indicates that even when the HD mode is not purely diffusive, it still has
the relation to the chaos.

(iv) Although the relation $D_{\mathrm{HD}}\sim v_{\mathrm{B}}^{2}\tau _{%
\mathrm{L}}$ is not accidental, it is not universal for all holographic
models. The limited universality that we have exhibited could be attributed
to both the existence of IR/UV fixed points and the simplicity of those
homogeneous holographic lattices. In the future, it is worth exploring
whether there is an explicit physical criterion that determines when $D_{%
\mathrm{HD}}\sim v_{\mathrm{B}}^{2}\tau _{\mathrm{L}}$ holds.

(v) For most holographic models in references, the bound $D_{\mathrm{T}%
}\gtrsim v_{\mathrm{B}}^{2}\tau _{\mathrm{L}}$ is saturated at low
temperatures or strong momentum relaxation. We have checked that this is
true for all the isotropic models that we have studied. However, the
violation of the bound $D_{\mathrm{T}}\gtrsim v_{\mathrm{B}}^{2}\tau _{%
\mathrm{L}}$ has been found in the inhomogeneous Sachdev-Ye-Kitaev chains
\cite{Gu1702} and quasi-topological Ricci polynomial gravities \cite{Lu1708}%
. Recently, by reconciling the conflict between the diffusive behavior and
operator growth lightcone, Hartman, Hartnoll and Mahajan proposed a new
bound $D\lesssim v^{2}\tau _{\mathrm{eq}}$ that is constituted by the
diffusivity $D$, equilibration timescale $\tau _{\mathrm{eq}}$ and lightcone
velocity $v$ \cite{Hartnoll1706}. For the holographic theories, $\tau _{%
\mathrm{eq}}$ can be determined by the leading non-hydrodynamic quasinormal
mode of black holes. The bound is obeyed in various weakly and strongly
interacting theories and can be relevant to the various transport. As an
example of this bound, it has been found that $D_{\mathrm{T}}\sim v_{\mathrm{%
B}}^{2}\tau _{\mathrm{eq}}$ in the EMA model with strong or weak momentum
relaxation. Note that the bound would be violated if $\tau _{\mathrm{eq}}$
is replaced by $\tau _{\mathrm{L}}$, since $\tau _{\mathrm{eq}}\gg \tau _{%
\mathrm{L}}$ when the momentum relaxation is weak. Let's compare $D_{\mathrm{%
T}}\sim v_{\mathrm{B}}^{2}\tau _{\mathrm{eq}}$ and $D_{\mathrm{HD}}\sim v_{%
\mathrm{B}}^{2}\tau _{\mathrm{L}}$. It can be understood that they indicate
two methods which might be useful to remedy the non-universal relation $D_{%
\mathrm{T}}\sim v_{\mathrm{B}}^{2}\tau _{\mathrm{L}}$. One is to change the
characteristic timescale and the other the diffusivity.

\section*{Acknowledgments}

We thank Andrea Amoretti, Elias Kiritsis, Hong Liu, Hong L\"{u}, Yan Liu,
Sang-Jin Sin, Zhuoyu Xian, and Jan Zaanen for helpful discussions. We
appreciate Blaise Gout\'{e}raux and Yi Ling for reading the manuscript and
providing valuable comments. We were supported partially by NSFC grants
(No.11675097, No.11575109, No.11375110, No.11475179, No. 11675015). Y.T. is
partially supported by the grants (No. 14DZ2260700) from Shanghai Key
Laboratory of High Temperature Superconductors. He is also partially
supported by the \textquotedblleft Strategic Priority Research Program of
the Chinese Academy of Sciences\textquotedblright , Grant No. XDB23030000. %
\appendix{} %\setcounter{equation}{0} \renewcommand{\theequation}{A.%
%\arabic{equation}}

\section{Decoupled thermo-electric currents}

Davison and Gout\'{e}raux have constructed two decoupled thermo-electric
currents in the EMA model \cite{DG1505}, which are correlated to the
decoupling of perturbation equations in the bulk. One of the currents is
expected to be transported by diffusion at all distance scales and the other
carries a gapped sound mode at short distances but it is diffusive at long
distances. Similar phenomenon has been found in the EMAD model \cite%
{Ling1512} but has not been studied in more general holographic states. Here
we will present a general expression of the decoupled currents which are
transported by diffusion at least at long distances.

Define two currents by the general linear combination of the electric
current and momentum%
\begin{equation}
J_{\pm }=J+\frac{1}{\gamma _{\pm }}P.  \label{Jzf}
\end{equation}%
These currents are decoupled $\left\langle J_{+}J_{-}\right\rangle =0$
provided that the coefficients $\gamma _{\pm }$ have the relation%
\begin{equation}
\gamma _{-}=-\frac{T\left( \alpha \gamma _{+}+\bar{\kappa}+2\alpha \mu
\right) +\mu \sigma \left( \gamma _{+}+\mu \right) }{T\alpha +\sigma \left(
\gamma _{+}+\mu \right) }.  \label{gammazf}
\end{equation}%
Simultaneously, the matrix of the conductivities can be diagonalized,%
\begin{equation}
\left(
\begin{array}{c}
J_{+} \\
J_{-}%
\end{array}%
\right) =\left(
\begin{array}{cc}
\frac{T\alpha +\sigma \left( \gamma _{+}+\mu \right) }{\gamma _{+}} & 0 \\
0 & \frac{T\left( T\alpha ^{2}-\bar{\kappa}\sigma \right) }{T\left( \alpha
\gamma _{+}+\bar{\kappa}+2\alpha \mu \right) +\mu \sigma \left( \gamma
_{+}+\mu \right) }%
\end{array}%
\right) \left(
\begin{array}{c}
-\nabla \mu +\left( \gamma _{-}+\mu \right) \nabla T/T \\
-\nabla \mu +\left( \gamma _{+}+\mu \right) \nabla T/T%
\end{array}%
\right) .
\end{equation}%
Next, we can require $J_{\pm }$ transported by diffusion at long distances,
that means%
\begin{equation}
J_{\pm }=-D_{\pm }\nabla \left( \delta \rho +\frac{1}{\gamma _{\pm }}\delta
\epsilon \right) .
\end{equation}%
Rigorously, this leads to%
\begin{equation}
\gamma _{+}=\frac{1}{2}\left[ \frac{\bar{\kappa}\chi -c_{\mu }\sigma }{\zeta
\sigma -\alpha \chi }-2\mu +\frac{1}{\zeta \sigma -\alpha \chi }\sqrt{\left(
\bar{\kappa}\chi -c_{\mu }\sigma \right) ^{2}-4T\left( c_{\mu }\alpha -\zeta
\bar{\kappa}\right) \left( \zeta \sigma -\alpha \chi \right) }\right] ,
\label{gammaz}
\end{equation}%
and the diffusion constants%
\begin{equation}
D_{\pm }=\frac{\chi \left[ \bar{\kappa}+\alpha \left( \gamma _{\pm }+\mu
\right) \right] -\zeta \left[ T\alpha +\sigma \left( \gamma _{\pm }+\mu
\right) \right] }{\left( c_{\mu }\chi -T\zeta ^{2}\right) }.  \label{Dzf}
\end{equation}%
Note that Eq. (\ref{Dzf}) denotes the eigenvalues of the diffusion matrix
which have been studied in \cite{Hartnoll1405}. As a consistent check of the
above derivation, one can find that the currents $J_{\pm }$ with special $%
\gamma _{\pm }$ are nothing but the decoupled currents found in the EMA
model \cite{DG1505} and the EMAD model \cite{Ling1512}.

We proceed to compare $J^{\mathrm{HD}}$ and $J_{\pm }$. Inserting the
diffusion condition $\bar{\kappa}\zeta =\alpha c_{\mu }$ into the decoupled
currents (\ref{Jzf}) with $\gamma _{-}$ (\ref{gammazf}) and $\gamma _{+}$ (%
\ref{gammaz}), one can prove $J^{\mathrm{HD}}=J_{+}$ (or $J_{-}$) due to $%
-\left( \bar{\kappa}+\alpha \mu \right) /\alpha =\gamma _{+}$ (or $\gamma
_{-}$) when $c_{\mu }\sigma \leq \bar{\kappa}\chi $ (or $c_{\mu }\sigma
\geqslant \bar{\kappa}\chi $). Thus, the HD current $J^{\mathrm{HD}}$ is
same as one of $J_{\pm }$ when they are transported by diffusion.

\section{Born-Infeld gravity with axions}

Here we will show a black-brane solution in the Born-Infeld gravity with the
axions and then calculate the dc thermo-electric conductivities. Consider
the background fields taking the form%
\begin{eqnarray}
ds^{2} &=&-f\left( r\right) dt^{2}+\frac{1}{f(r)}dr^{2}+r^{2}\left(
dx_{1}^{2}+dx_{2}^{2}\right) ,  \notag \\
A_{t} &=&A_{t}(r),\;\chi _{i}=\beta x_{i}.  \label{BIansatz2}
\end{eqnarray}%
Variation of the action (\ref{BIaction}) with respect to $A_{t}(r)$
generates the Maxwell equation%
\begin{equation}
A_{t}^{\prime \prime }+\frac{2}{r}A_{t}^{\prime }\left( 1-2\beta
A_{t}^{\prime 2}\right) =0.
\end{equation}%
Its integration leads to%
\begin{equation}
A_{t}^{\prime }=\frac{1}{Z_{eff}(r)}\frac{\rho }{r^{2}},\;\text{with }%
Z_{eff}(r)=\sqrt{1+\frac{b\rho ^{2}}{r^{4}}}.  \label{BIAt}
\end{equation}%
Inserting Eq. (\ref{BIansatz2}) and Eq. (\ref{BIAt}) into the Einstein
equations, one can find the only nontrivial component
\begin{equation}
f^{\prime }+\frac{f}{r}+\frac{\beta ^{2}}{2r}+\frac{r}{2b}\left[
Z_{eff}(r)-1-6b\right] =0.
\end{equation}%
The solution is%
\begin{equation}
f=r^{2}-\frac{\beta ^{2}}{2}-\frac{m}{r}+\frac{r^{2}}{6b}\left[
1-Z_{eff}\left( r\right) \right] +\frac{\rho ^{2}}{3r^{2}}\;_{2}F_{1}\left[
\frac{1}{4},\frac{1}{2},\frac{5}{4},-\frac{b\rho ^{2}}{r^{4}}\right] ,
\label{blacknessBI}
\end{equation}%
where $_{2}F_{1}$ is a hypergeometric function and $m$ is the mass parameter.

We will turn to calculate the dc thermo-electric conductivities using the
Donos-Gauntlett method \cite{Donos1406}. The consistent ansatz of the
perturbation along $x=x_{1}$ reads%
\begin{eqnarray}
A_{x} &=&-Et+\xi A_{t}t+a_{x}(r),\;g_{tx}=-\xi tf(r)+r^{2}h_{tx}(r),  \notag
\\
g_{rx} &=&r^{2}h_{rx}(r),\;\chi _{1}=\beta x+\psi _{1}(r).
\end{eqnarray}%
Note that the former two modes have linear terms depending on time. This
ansatz corresponds to applying an external electric field $E$ and
temperature gradient $\nabla T=\xi T$ to the boundary theory. The other
metric components and fields have the perturbation depending only on the
radial coordinate.

In order to determine the dc conductivities, one needs to impose the
regularity of fluctuation modes on the horizon:%
\begin{eqnarray}
a_{x} &=&-\frac{E}{f}+\mathcal{O}(r-r_{+}),  \notag \\
h_{tx} &=&fh_{rx}-\frac{\xi f}{r^{2}}\int \frac{1}{f}dr+\mathcal{O}(r-r_{+}).
\end{eqnarray}%
The $rx$ component of the linearized Einstein equations gives%
\begin{equation}
h_{rx}=\frac{-\rho E}{\beta ^{2}r^{2}f}+\frac{\xi }{\beta ^{2}}\left( \frac{2%
}{r}+\frac{\rho A_{t}}{r^{2}f}-\frac{h^{\prime }}{h}\right) +\frac{\psi
_{1}^{\prime }}{\beta }.
\end{equation}%
It is important to construct two conserved currents which are independent of
the radial coordinate. The first one can be obtained from the Maxwell
equations:%
\begin{equation}
J^{x}=-\sqrt{-g}Z_{eff}F^{rx}=-\rho h_{tx}-Z_{eff}fa_{x}^{\prime }.
\end{equation}%
The second one is built up through introducing a two-form associated with
the Killing vector $k=\partial _{t}$:%
\begin{equation}
G^{\mu \nu }=2\nabla ^{\mu }k^{\nu }+Z_{eff}k^{[\mu }F^{\nu ]\sigma
}A_{\sigma }+\frac{1}{2}\left( \psi -2\theta \right) Z_{eff}F^{\mu \nu },
\end{equation}%
where%
\begin{equation}
\psi =-Ex,\;\theta =-Ex-A_{t}.
\end{equation}%
We only concern its $rx$ component%
\begin{eqnarray}
Q^{x} &=&\sqrt{-\gamma }G^{rx}  \notag \\
&=&\left[ \rho A_{t}+\left( 2rh-r^{2}f^{\prime }\right) \right]
h_{tx}+f\left( Z_{eff}A_{t}a_{x}^{\prime }+r^{2}h_{tx}^{\prime }\right) .
\end{eqnarray}

Consider the renormalized on-shell action \cite{Tan0903}%
\begin{equation}
S_{\mathrm{os}}=\left( S_{\mathrm{bulk}}+S_{\mathrm{GH}}+S_{\mathrm{ct}%
}\right) _{\mathrm{on-shell}},
\end{equation}%
where the Gibbs-Hawking term%
\begin{equation}
S_{\mathrm{GH}}=-2\int d^{3}x\sqrt{-\tilde{\gamma}}K
\end{equation}%
is supplied to implement a well-defined variational principle and the
counterterms%
\begin{equation}
S_{\mathrm{ct}}=\int d^{3}x\sqrt{-\tilde{\gamma}}\left( -4+\frac{1}{2}%
\sum\limits_{i=1}^{2}\nabla _{a}\chi _{i}\nabla ^{a}\chi _{i}\right)
\end{equation}%
are invoked to cancel the UV divergence. Here $\tilde{\gamma}_{ab}$ is the
induced metric on the boundary, $\tilde{\gamma}$ is its determinant and $K$
the trace of the extrinsic curvature. The covariant currents $\tilde{T}^{\mu
\nu }$ and $\tilde{J}^{\mu }$ can then be calculated by%
\begin{eqnarray}
\tilde{T}^{ab} &=&\frac{2}{\sqrt{-\tilde{\gamma}}}\frac{\delta S_{\mathrm{os}%
}}{\delta \tilde{\gamma}_{ab}}  \notag \\
&=&2\left[ K^{ab}-K\tilde{\gamma}^{ab}-2\tilde{\gamma}^{ab}-\frac{1}{2}%
\sum\limits_{i=1}^{2}\left( \nabla ^{a}\chi _{i}\nabla ^{b}\chi _{i}+\tilde{%
\gamma}^{ab}\nabla _{c}\chi _{i}\nabla ^{c}\chi _{i}\right) \right] ,  \notag
\\
\tilde{J}^{a} &=&\frac{1}{\sqrt{-\tilde{\gamma}}}\frac{\delta S_{\mathrm{os}}%
}{\delta A_{a}}=-n_{b}F^{ba}\left( 1+\frac{b}{2}F^{\mu \nu }F_{\mu \nu
}\right) ^{-1/2},
\end{eqnarray}%
where $n_{b}$ is the normal vector of the boundary. One can prove that $%
J^{x} $ matches the electric current $\sqrt{-\tilde{\gamma}}\tilde{J}^{x}$
on the AdS boundary exactly. $Q^{x}$ also matches the thermal current $\sqrt{%
-\tilde{\gamma}}\left( \tilde{\gamma}_{xx}\tilde{T}^{tx}-\mu \tilde{J}%
^{x}\right) $, up to a term depending on the time linearly. But this term
does not contribute to the dc conductivity \cite{Donos1406}. Thus, by
evaluating $J^{x}$ and $Q^{x}$ on the horizon, one can extract the dc
conductivities%
\begin{equation}
\sigma =\frac{\partial J^{x}}{\partial E}=Z_{eff}+\frac{4\pi \rho ^{2}}{%
\beta ^{2}s},\;\alpha =\frac{1}{T}\frac{\partial J^{x}}{\partial \xi }=\frac{%
4\pi \rho }{\beta ^{2}},\;\bar{\kappa}=\frac{1}{T}\frac{\partial Q^{x}}{%
\partial \xi }=\frac{4\pi sT}{\beta ^{2}}.
\end{equation}

\section{A counter example}

We will study an EMAD theory with the action%
\begin{equation}
S_{\mathrm{bulk}}=\int d^{4}x\sqrt{-g}\left[ R+V\left( \phi \right) -\frac{1%
}{2}\left( \partial \phi \right) ^{2}-\frac{1}{4}Z\left( \phi \right) F^{2}-%
\frac{1}{2}\sum\limits_{i=1}^{2}\left( \partial \chi _{i}\right) ^{2}\right]
,
\end{equation}%
where the gauge field coupling is $Z\left( \phi \right) =e^{-\delta \phi }$
and the potential of the dilaton involves three different exponential
functions%
\begin{equation}
V\left( \phi \right) =V_{1}e^{\frac{\delta ^{2}-1}{2\delta }}+V_{2}e^{-\frac{%
1}{\delta }}+V_{3}e^{\delta },
\end{equation}%
with the parameters $\delta $ and%
\begin{equation}
V_{1}=-\frac{16\delta ^{2}}{\left( \delta ^{2}+1\right) ^{2}},\;V_{2}=\frac{%
2\delta ^{2}\left( 1-3\delta ^{2}\right) }{\left( \delta ^{2}+1\right) ^{2}}%
,\;V_{3}=\frac{2\left( \delta ^{2}-3\right) }{\left( \delta ^{2}+1\right)
^{2}}.
\end{equation}%
Setting $\chi _{i}$ as two linear axions, an analytical black-brane solution
has been found in this theory \cite{Gouteraux1401}. We write down the line
element%
\begin{equation}
ds^{2}=-f(r)h(r)^{\frac{-2}{\delta ^{2}+1}}dt^{2}+h(r)^{\frac{2}{\delta
^{2}+1}}\left[ \frac{dr^{2}}{f(r)}+r^{2}\left( dx^{2}+dy^{2}\right) \right] ,
\end{equation}%
where%
\begin{eqnarray}
f(r) &=&r^{2}\left[ h(r)^{\frac{4}{\delta ^{2}+1}}-\left( \frac{r_{+}}{r}%
\right) ^{3}h(r_{+})^{\frac{4}{\delta ^{2}+1}}\right] -\frac{\beta ^{2}}{2}%
\left( 1-\frac{r_{+}}{r}\right) ,  \notag \\
h(r) &=&1+\frac{Q}{r}.
\end{eqnarray}%
The parameter $Q$ is related to the charge density $\rho $ by%
\begin{equation}
\rho ^{2}=\frac{2Q}{\delta ^{2}+1}\left[ 2r_{+}^{3}\left( \frac{Q+r_{+}}{%
r_{+}}\right) ^{\frac{4}{\delta ^{2}+1}}-\beta ^{2}\left( Q+r_{+}\right) %
\right] .
\end{equation}%
The Hawking temperature and entropy density can be read:%
\begin{equation}
T=-\frac{h(r_{+})^{\frac{-2}{\delta ^{2}+1}}}{4\pi r_{+}}\left\{ \frac{\beta
^{2}}{2}+\frac{r_{+}^{2}}{\delta ^{2}+1}h(r_{+})^{\frac{4}{\delta ^{2}+1}}%
\left[ 1-3\delta ^{2}-\frac{4}{h(r_{+})}\right] \right\} ,\;s=4\pi h(r_{+})^{%
\frac{2}{\delta ^{2}+1}}r_{+}^{2}.  \label{TS EMDA}
\end{equation}%
The thermo-electric conductivities have been obtained in Ref. \cite%
{Donos1406}
\begin{equation}
\sigma =Z+\frac{4\pi \rho ^{2}}{s\beta ^{2}},\;\alpha =\frac{4\pi \rho }{%
\beta ^{2}},\;\bar{\kappa}=\frac{4\pi sT}{\beta ^{2}}.  \label{sakEMDA}
\end{equation}%
In terms of Ref. \cite{Lu1701}, the chaos quantities can be expressed as
\begin{equation}
\tau _{\mathrm{L}}=\frac{1}{2\pi T},\;v_{\mathrm{B}}^{2}=\left( \frac{Q+r_{+}%
}{r_{+}}\right) ^{1-\frac{2}{\delta ^{2}+1}}\frac{\delta ^{2}+1}{2\left[
r_{+}+\left( Q+r_{+}\right) \delta ^{2}\right] }.  \label{VBEMAD}
\end{equation}%
Putting Eqs. (\ref{TS EMDA}), (\ref{sakEMDA}) and (\ref{VBEMAD}) together,
we can calculate the ratio between $D_{\mathrm{HD}}$ and $v_{\mathrm{B}%
}^{2}\tau _{\mathrm{L}}$, giving%
\begin{equation}
\frac{D_{\mathrm{HD}}}{v_{\mathrm{B}}^{2}\tau _{\mathrm{L}}}=\frac{%
r_{+}+\left( Q+r_{+}\right) \delta ^{2}}{\left( Q+r_{+}\right) \left( \delta
^{2}+1\right) }\frac{\beta ^{2}\left( Q+r_{+}\right) \left( \delta
^{2}+1\right) \!-\!2r_{+}^{2}\left( \frac{Q+r_{+}}{r_{+}}\right) ^{\frac{4}{%
\delta ^{2}+1}}\left( Q\!+\!r_{+}\!-\!3Q\delta ^{2}\!+\!r_{+}\delta
^{2}\right) }{\beta ^{2}\left[ r_{+}+\left( 2Q+r_{+}\right) \delta ^{2}%
\right] -2r_{+}^{3}\left( \frac{Q+r_{+}}{r_{+}}\right) ^{\frac{4}{\delta
^{2}+1}}\left( \delta ^{2}+1\right) },
\end{equation}%
which does not equal to one for any $\delta \neq 0$. Keeping in mind that
the near-horizon geometry of these black branes is $AdS_{2}\times R^{2}$ for
$\delta \neq \sqrt{1/3}$\footnote{%
For $\delta =\sqrt{1/3}$, it can be conformal to $AdS_{2}$ \cite%
{Gouteraux1401}. Note that the Kelvin formula is not obeyed in this case.},
we can know that the existence of the IR/UV fixed points is not sufficient
to ensure the relation $D_{\mathrm{HD}}\sim v_{\mathrm{B}}^{2}\tau _{\mathrm{%
L}}$.

\end{document}